\documentclass[english,twocolumn,prd,showpacs]{revtex4}
\usepackage{graphicx}

\newcommand{\beqa}{\begin{eqnarray*}}\newcommand{\eeqa}{\end{eqnarray*}}
\newcommand{\beq}{\begin{eqnarray}}\newcommand{\eeq}{\end{eqnarray}}
\newcommand{\bwt}{\begin{widetext}}\newcommand{\ewt}{\end{widetext}}

\renewcommand{\a}{\alpha}\newcommand{\as}{\alpha_s}
\renewcommand{\b}{\beta}
\newcommand{\de}{\delta_\e}
\newcommand{\g}{\gamma}
\newcommand{\s}{\sigma}
\newcommand{\e}{\epsilon}\newcommand{\ve}{\varepsilon}

\newcommand{\MeV}{\,\text{MeV}}\newcommand{\GeV}{\,\text{GeV}}
\newcommand{\fb}{\,\text{fb}}

\newcommand{\epem}{e^+e^-}\newcommand{\JP}{J/\psi}\newcommand{\ec}{\eta_c}
\newcommand{\OJ}{O_\psi}\newcommand{\Oe}{O_\eta}
\newcommand{\M}{{\cal M}}

\newcommand{\arctanh}{\text{arctanh}\,}
\newcommand{\res}{{\text{res}\,}}
\newcommand{\hp}{\hat p}\newcommand{\hs}{\hat s}
\newcommand{\bv}{\bar v\left(k_2\right)}
\renewcommand{\u}{u\left(k_1\right)}

\begin{document}
\title{Initial state radiation corrections to double charmonium production in one-photon electron-positron annihilation}
\author{A.V. Luchinsky}
\affiliation{Institute for High Energy Physics, Protvino, Russia}
\email{Alexei.Luchinsky@ihep.mail.ru}
\begin{abstract}
In this paper initial state radiation corrections to double production of charmonium mesons on one-photon electron-positron annihilation at center of mass energy $\sqrt{s}=10.6\GeV$ are studied. It is shown, that these corrections have noticeable effect and must be taken into consideration.
\end{abstract}
\date{\today}
\maketitle

\section{Introduction\label{Introduction}}
Since the discovery of the so called charmonium mesons (that is mesons consisting of $c$- and $\bar c$-quarks, for example $J/\psi$, $\psi(2S)$ or $\eta_c$) in 1974 \cite{Aubert:1974js,Augustin:1974xw} they play a significant role in our understanding of quantum chromodynamics (QCD). Production of charmonium requires creation of heavy $c\bar c$ pair with the energy greater than $2m_c$ where QCD coupling constant is small enough to use perturbation theory. However, the subsequent hadronization probes much smaller mass scales of order $m_c v^2$, where $v$ is typical velocity of $c$ quark in the charmonium rest frame. For $J/\psi$, $m_c v^2$ is numerically of order $\Lambda_{\rm QCD}$, so the production processes are sensitive to nonperturbative physics as well. Charmonium mesons are also very interesting from the experimental point of view, since the narrow resonances can easily be separated from the background.

In 1986 Caswell and Lepage have proposed the Non-Relativistic Quantum Chromodynamics model (NRQCD) \cite{Caswell:1986ui,Bodwin:1995jh}. This model exploits the fact that the velocity $v$ of the heavy quark in the charmonium rest frame is small in comparison with the speed of light and all physical quantities can be expressed as the series in this small velocity and electromagnetic and strong coupling constants $\a$ and $\as$. Matrix elements of the four-fermion operators, used in this theory, can be determined phenomenologically. Both inclusive and exclusive production of charmonium mesons and their decays were studied using this theory \cite{Braaten:1996ez,Cho:1996cg,Bodwin:2002aa,Braaten:2002aa,Bodwin:2003aa,Hagivara:2003aa,Wang:2002aa}.

New difficulties have arisen as a result of recent measurements of inclusive charmonium production in $\epem$ collisions by Belle and BABAR collaborations  \cite{Aubert:2001pd,Abe:2001aa}. Their results were about an order of magnitude higher, than the predictions based on NRQCD or other models \cite{Kiselev:1994pu,Braaten:1996ez,Berezhnoy:2003hz,Leibovich:2003aa,Luchinsky:2003ej}.

Another difference has arisen in studying the exclusive production of pair of charmonium mesons. For example, the NRQCD results for $J/\psi\eta_c$ production is \cite{Braaten:2002aa}
\beqa
\s_0&=&\s\left(e^+e^-\to\g^*\to J/\psi\eta_c\right)=2.31\fb
\eeqa
and Belle results \cite{Abe:2002aa} are an order of magnitude higher. One of the possible explanations of this difference was proposed in \cite{Bodwin:2002fk}. The authors assumed that some of the Belle's $J/\psi\eta_c$ signals could actually be the double production of $J/\psi$ meson with subsequent decay $J/\psi\to\eta_c\g$ and presented the value
\beqa
\s\left(\epem\to2\g^*\to2J/\psi\right)&=&8.7\fb.
\eeqa
Later in \cite{Braaten:2002aa,Luchinsky:2003yh} it was shown that due to relativistic and higher order QCD corrections the cross section should be about 4 times smaller and the question of the difference between theory and experiment remains open.

In this paper I propose that this difference can be explained by the initial state radiation corrections. Indeed, the naive estimate of the effect of these corrections gives us the result
\beqa
\s&=&\s\left(\epem\to J/\psi\eta_c\g\right)\sim
\a\log\frac{s}{m_e^2}\s_0
\eeqa
and the suppression caused by additional factor $\a$ will be compensated by large logarithm $\log{s/m_e^2}$. Another reason is that cross section of the reaction $\epem\to\JP\ec$ as a function of center of mass energy $\sqrt{s}$ has a narrow peak near $\sqrt{s}\approx6.5\,\GeV$, where $\s\left(\epem\to\JP\ec\right)\approx30\fb$ and by emitting a hard photon we can return to this region.

The rest of this paper is organized as follows. In the next section I give the diagrams and matrix element for the process under consideration. In section \ref{Total} the method of calculation of the total cross section is described. Finally, distributions over vector charmonium energy and scattering angle are presented.

\section{Matrix element\label{Matrix}}
\begin{figure*}
\includegraphics*{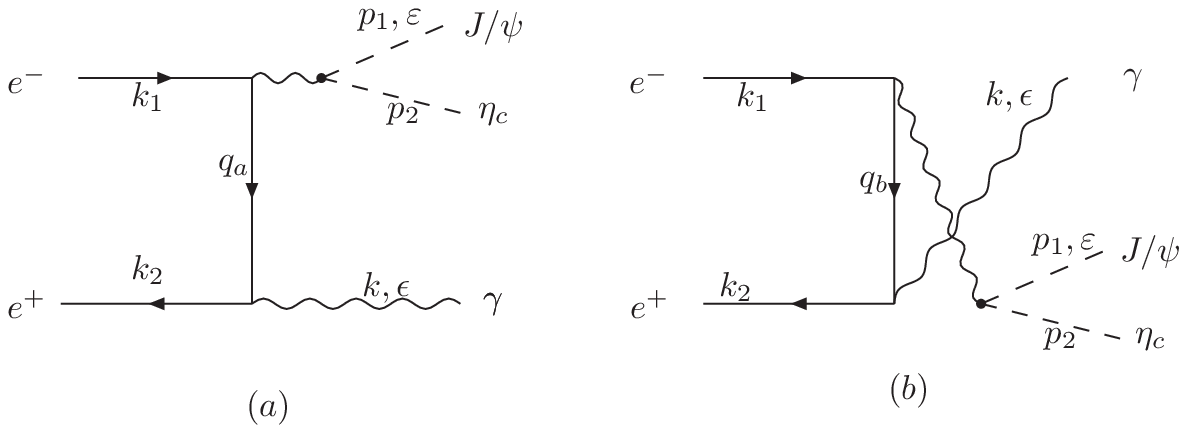}
\caption{Diagrams for $\epem\to\JP\ec\g$.}\label{Diag1}
\end{figure*}

\begin{figure*}
\includegraphics*{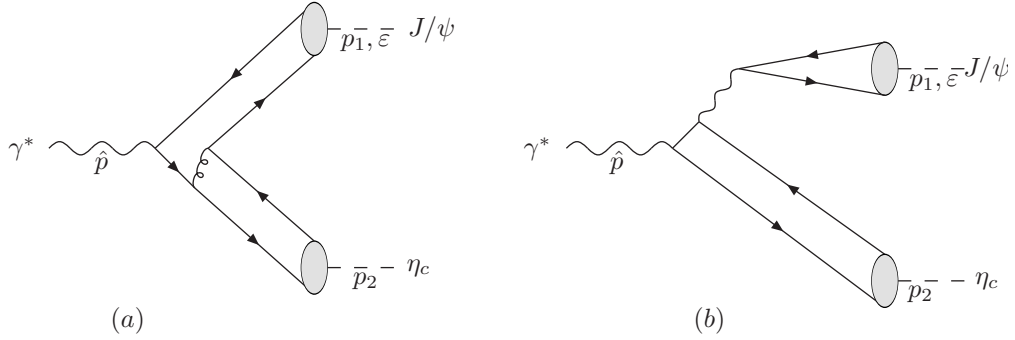}
\caption{Some diagrams for the transition of a virtual photon into charmonium pair. Other diagrams can be obtained from theses by permutations.}\label{Diag2}
\end{figure*}

The diagrams for the process $e^-\left(k_1\right)e^+\left(k_2\right)\to\JP\left(p_1,E\right)\ec\left(p_2\right)\g\left(k,\e\right)$ are shown on figure \ref{Diag1}. The vertex  of the transition of virtual photon with momentum $\hp$ and its square $\hs=\hp^2$ into pair of charmonium states, denoted here by filled circle, was calculated in \cite{Braaten:2002aa}. Some of the diagrams of this transition are shown on figure \ref{Diag2}, the others can be obtained from them by permutations. One can also add electromagnetic diagrams by replacing the gluon line in figure \ref{Diag2}(a) by photon, but they are suppressed by the factor $\a/\as$ and will not be considered here. On the other hand, as it was shown in \cite{Braaten:2002aa}  the purely electromagnetic resonance diagrams (for example, the diagram \ref{Diag2}(b)) must be included. The reason for this is that the additional factor $\a$ is compensated by $r^2=4M^2/\hs$, where $M=2m_c$ is the mass of the charmonium meson (the difference between $\JP$ and $\ec$ masses is neglected) and $m_c$ is $c$-quark mass.

The analytical expression for the $\g\JP\ec$ vertex was found to be
\beqa
J_\mu&=&\left<\JP\left(p_1,\ve\right)\ec\left(p_2\right)
\left|\bar c\g_\mu c\right|
0\right>=\\
&=&i A\e_{\mu\nu\a\b}\ve^\nu p_1^\a p_2^\b,
\eeqa
where $\ve$ is the polarization vector of the vector charmonium $\JP$,  the coefficient $A$ is
\beqa
A&=&\frac{128\pi\as}{\hs^2}\sqrt{\OJ\Oe}
\left(
\frac{N_c^2-1}{2N_c^2}+\frac{e_c^2\a}{N_c\as}-\frac{1}{r^2}\frac{e_c^2\a}{\as}
\right),
\eeqa
$N_c=3$ is the number of colors, $e_c=2/3$ is the charge of the $c$ quark (in the units of the elementary charge $e$), and the constants
\beqa
\OJ&=&\left<O_1\right>_{\JP},\qquad\Oe=\left<O_1\right>_{\ec}
\eeqa
are the matrix elements of the NRQCD probability factors. These constants can be determined from the experimental data on $\JP$ and $\ec$ decay widths \cite{Braaten:2002aa}:
\beqa
\OJ&=&
\frac{27}{32\pi\a^2}\Gamma\left(\JP\to\epem\right)M^2\left(1-\frac{8}{3}\frac{\as}{\pi}\right)^{-2},\\
\Oe&=&
\frac{81}{128\pi\a^2}\Gamma\left(\ec\to\g\g\right)
\left(1-\frac{2-\pi^2}{6}\frac{\as}{\pi}\right)^{-2}.
\eeqa
Using the table data on particle decay widths \cite{Groom:2000in} we get the numerical values of probability factors, that will be used in this paper:
\beqa
\OJ&=&0.335\GeV^3,\\
\Oe&=&0.297\GeV^3.
\eeqa

The matrix element corresponding to the diagrams shown on figure \ref{Diag1} is
\beq
\M&=&\frac{e_c e^3}{\hs}\e^\mu J^\nu \bv\left\{	
\frac{1}{q_a^2}\g_\nu\hat q_a\g_\mu+\right.\nonumber\\
&&\left.\frac{1}{q_b^2}\g_\mu\hat q_b\g_\nu\right\}\u,\label{Matr}
\eeq
where $k_1$ and $k_2$ are electron and positron momenta (the mass of the electron $m_e$ is neglected wherever it is possible), $k$ is the momentum of the final state photon, $\e_\mu$ is its polarization vector, $q_a=k_1-k$ and $q_b=k-k_2$ are the momentum transfer in this two diagrams. The cross section found from (\ref{Matr})  and equals to
\beqa
\s&=&\frac{1}{2048\pi^7\left(k_1k_2\right)}
\int d\Phi_3\left(k_1+k_2;p_1,p_2,k\right)\sum\left|\M\right|^2,
\eeqa
where summation is performed over polarizations of initial and final particles and for the Lorentz-invariant phase space  the notation
\beq
d\Phi_n\left(P;p_1,\ldots,p_n\right)&=&
\delta\left(P-\sum\limits_{i=1}^n p_i\right)\prod\limits_{i=1}^n\frac{d^3p_i}{2E_i} \label{LIPS}
\eeq
is used.

\section{Total cross section\label{Total}}

The Lorentz-invariant phase space (\ref{LIPS}) satisfies the recursion relation
\beqa
d\Phi_3\left(k_1+k_2;p_1,p_2,k\right)&=& d\hs
d\Phi_2\left(k_1+k_2;\hp,k\right)\\
&\times&d\Phi_2\left(\hs;p_1,p_2\right),
\eeqa
where $\hp=p_1+p_2$ is the momentum of the virtual photon, $\hs=\hp^2=s-2\sqrt{s}E_k$, and $E_k$ is the final state photon energy in the laboratory frame. Leaving only the integration over $E_k$ we get the differential cross section
\beqa
\frac{d\s}{d\hs}&=&\frac{32768\pi^2\a^3}{19683}\frac{\OJ\Oe}{s^2}\left(1-\frac{4M^2}{\hs}\right)^{3/2}\times\\
&\times&\frac{(2\a\hs-4M^2(\a+2\as))^2}{\hs^4}\times\\
&\times&\left(2\frac{s^2+\hs^2}{s-\hs}\frac{\arctanh\beta_e}{\beta_e}-s+\hs\right).
\eeqa
From this equation one can easily notice two important items:
\begin{itemize}
\item the cross section is enhanced by the factor $\arctanh\beta_e\sim\log s/m_e^2$ as it was states in the introduction,
\item total cross section actually diverges in the limit $\hs\approx s$. This fact is the familiar infrared divergence caused by the emission of the soft photon, where one can not use the perturbation theory. To avoid this divergence one must put the cutoff $\de$ on the energy of the photon:
\beqa
E_k&>&\de.
\eeqa
\end{itemize}

\begin{figure}
\includegraphics[width=8cm]{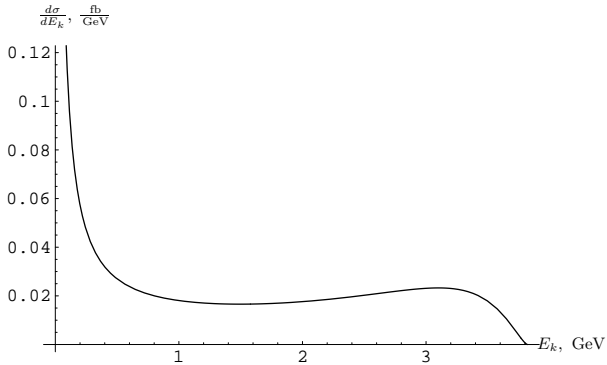}
\caption{Distribution over final state photon energy}\label{DsDw}
\end{figure}

The plot for differential cross section $d\s/dE_k$ is presented on figure \ref{DsDw}. As stated above, in the limit of small photon energy it grows to infinity and one should put the cutoff. The bump in the region $E_k\approx 3\GeV$ corresponds to $\sqrt{\hs}\approx6.5\GeV$, where the cross section of the process $\g^*\to\JP\ec$ is maximal.

Values of total cross section for some cutoff photon energies are:
\beqa
\s&\approx&2.84\fb\quad\rm{ for\ }\de=1\MeV,\\
\s&\approx&2.234\fb\quad\rm{ for\ }\de=10\MeV,\\
\s&\approx&1.72\fb\quad\rm{ for\ }\de=100\MeV.
\eeqa
It is easily seen, that by the order of magnitude these cross sections equals to the cross section of the base process $\epem\to\g^*\to\JP\ec$, so the initial state radiation corrections should be taken into account.

\section{Differential cross section\label{Differential}}

For the calculation of the differential cross section over the vector charmonium energy it is convenient to introduce dimensionless variables according to formula's
\beqa
z_{k,1,2}&=&\frac{2E_{k,1,2}}{\sqrt s},\\
\mu&=&\frac{2M}{\sqrt s},\\
x_{k,1,2}&=&\cos\theta_{k,1,2},\\
y_{k,1,2}&=&\sqrt{1-x_{k,1,2}^2}\,,
\eeqa
where $E_1$, $E_2$ and $E_k$ are the energies of vector and pseudoscalar charmonia and photon in the center of mass frame respectively; $\theta_1$, $\theta_2$ and $\theta_k$ are the angles between the momentum of the initial electron and the momenta of these final particles. These variables take the values in the regions
\beqa
x_1,x_k	&\in&	[-1;1],\\
z_1    	&\in&	[\mu;1],\\
z_k			&\in&	[z_k^m;z_k^M],
\eeqa
where minimum and maximum values of $z_k$ are
\beqa
z_k^{m,M}&=&\frac{2(1-z_1)}{2-z_1[1-\b(x_1x_k\pm y_1y_k)]}.
\eeqa
In the terms of these variables the squared matrix element and Lorentz-invariant phase space take the form
\bwt\beqa
\left|\M\right|^2&=&\frac{13417728\a^3}{6561\mu^4}\frac{\OJ\Oe}{s^4}
 \frac{(3\as\mu-\a(3-3z_k-\mu^2))^2}{z_k^2(1-z_k)^5(1-\b_e^2x_k^2)}\left\{\rule{0cm}{.5cm}
 	4\left(2-\mu^2-2z_1+(1+\b^2x_1^2)z_1^2\right)- \right.
\\&&
	4\left(4-\mu^2-3\left(3+\b x_1 x_k\right)z_1+\left(1+\b^2x_1^2\right)z_1^2\right)z_k+
\\&&
	\left(2\left(5-2z_1\left(1-\b x_1 x_k\right)+x_k^2\right)-\mu^2\left(1+x_k^2\right)\right)z_k^2-
\\&&\left.\rule{0cm}{.5cm}2\left(1+x_k^2\right)z_k^3\right\},\\
d\Phi_3\left(k_1+k_2;p_1,p_2,k\right)&=&
\frac{\pi s}{16}\frac{\b z_1 z_k}{1-z_1}\sqrt{\frac{z_k^M z_k^m}{(z_k^M-z_k)(z_k-z_k^m)}}
dz_1 dx_1 dz_k dx_k,
\eeqa\ewt
where $\b=\sqrt{1-M^2/E_1^2}=\sqrt{1-\mu^2/z_1^2}$ is the velocity of the vector charmonium meson and $\b_e=\sqrt{1-4m_e^2/s}$ is the velocity of the initial electron in the laboratory frame (this is the only place where I have to leave the non-vanishing electron mass).

For integration of the differential cross section over $z_k$ it is possible to use the residue technique. The only $d^4\s/dx_1 dz_1 dx_z dz_k$ poles are $z_k=0,1$ in , so we can write
\beqa
\frac{d^3\s}{dx_1 dz_1 dx_k}&=&i\pi\sum\limits_{z_k=0,1}\res\frac{d^4\s}{dx_1 dz_1 dx_k dz_k}.
\eeqa
The residue at $z_k=1$ is too lengthy to be presented here, but the residue at $z_k=0$, that plays most important role for $z_1\approx1$ (that is in the limit of maximal vector charmonium energy) and $x_1\ne\pm1$ it is rather compact. In this limit we have
\beqa
\frac{d^2\s}{dz_1dx_1}&=&S(z_1)\left[1+B(z_1)x_1^2\right],\\
S(z_1)&=&\frac{262144\pi^2\a^3\b}{6561\mu^4}\frac{\OJ\Oe}{s^4}\frac{\arctanh\b_e}{\b_e}z_1\times\\
&\times&\frac{(2-\mu^2-2z_1+z_1^2)(3\as\mu^2-\a(3-\mu^2))^2}{1-z_1},\\
B(z_1)&=&\frac{\b^2z_1^2}{2-\mu^2-2z_1+z_1^2}.
\eeqa
The integration over all other variables was performed numerically.

\begin{figure}
\includegraphics[width=8cm]{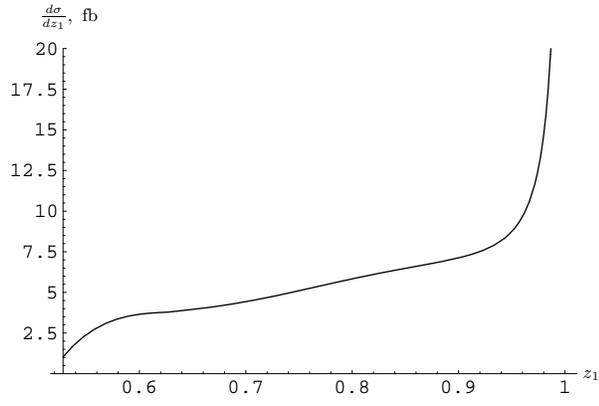}
\caption{Distribution over vector charmonium energy.}\label{DsDz1}
\end{figure}

The distribution over vector charmonium energy is shown on figure \ref{DsDz1}. In the limit of maximal vector charmonium energy ($z_1\approx1$) this curve grows to infinity corresponding to analogous grows int the distribution over photon energy shown on figure \ref{DsDw}.

\begin{figure}
\includegraphics[width=8cm]{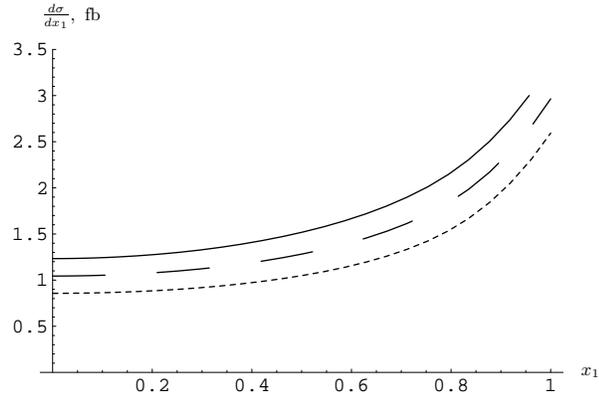}
\caption{Distribution over scattering angle for $\de=1\MeV$ (solid line), $\de=10\MeV$ (dashed line), and $\de=100\MeV$ (dotted line)}\label{DsDx1}
\end{figure}

The distributions over scattering angle for different values of $\de$ are shown on figure \ref{DsDx1}. This form of distribution is usual for color-singlet charmonium production.

\section*{Acknowledgments}
Author would like to thank A.K. Likhoded for useful discussions. This work was partly supported by Russian Scientific School, grant \#1202.2003.2.


\begin{thebibliography}{21}
\expandafter\ifx\csname natexlab\endcsname\relax\def\natexlab#1{#1}\fi
\expandafter\ifx\csname bibnamefont\endcsname\relax
  \def\bibnamefont#1{#1}\fi
\expandafter\ifx\csname bibfnamefont\endcsname\relax
  \def\bibfnamefont#1{#1}\fi
\expandafter\ifx\csname citenamefont\endcsname\relax
  \def\citenamefont#1{#1}\fi
\expandafter\ifx\csname url\endcsname\relax
  \def\url#1{\texttt{#1}}\fi
\expandafter\ifx\csname urlprefix\endcsname\relax\def\urlprefix{URL }\fi
\providecommand{\bibinfo}[2]{#2}
\providecommand{\eprint}[2][]{\url{#2}}

\bibitem[{\citenamefont{Aubert et~al.}(1974)}]{Aubert:1974js}
\bibinfo{author}{\bibfnamefont{J.~J.} \bibnamefont{Aubert}}
  \bibnamefont{et~al.}, \bibinfo{journal}{Phys. Rev. Lett.}
  \textbf{\bibinfo{volume}{33}}, \bibinfo{pages}{1404} (\bibinfo{year}{1974}).

\bibitem[{\citenamefont{Augustin et~al.}(1974)}]{Augustin:1974xw}
\bibinfo{author}{\bibfnamefont{J.~E.} \bibnamefont{Augustin}}
  \bibnamefont{et~al.}, \bibinfo{journal}{Phys. Rev. Lett.}
  \textbf{\bibinfo{volume}{33}}, \bibinfo{pages}{1406} (\bibinfo{year}{1974}).

\bibitem[{\citenamefont{Caswell and Lepage}(1986)}]{Caswell:1986ui}
\bibinfo{author}{\bibfnamefont{W.~E.} \bibnamefont{Caswell}} \bibnamefont{and}
  \bibinfo{author}{\bibfnamefont{G.~P.} \bibnamefont{Lepage}},
  \bibinfo{journal}{Phys. Lett.} \textbf{\bibinfo{volume}{B167}},
  \bibinfo{pages}{437} (\bibinfo{year}{1986}).

\bibitem[{\citenamefont{Bodwin et~al.}(1995)\citenamefont{Bodwin, Braaten, and
  Lepage}}]{Bodwin:1995jh}
\bibinfo{author}{\bibfnamefont{G.~T.} \bibnamefont{Bodwin}},
  \bibinfo{author}{\bibfnamefont{E.}~\bibnamefont{Braaten}}, \bibnamefont{and}
  \bibinfo{author}{\bibfnamefont{G.~P.} \bibnamefont{Lepage}},
  \bibinfo{journal}{Phys. Rev.} \textbf{\bibinfo{volume}{D51}},
  \bibinfo{pages}{1125} (\bibinfo{year}{1995}), \eprint{hep-ph/9407339}.

\bibitem[{\citenamefont{Braaten and Chen}(1996)}]{Braaten:1996ez}
\bibinfo{author}{\bibfnamefont{E.}~\bibnamefont{Braaten}} \bibnamefont{and}
  \bibinfo{author}{\bibfnamefont{Y.-Q.} \bibnamefont{Chen}},
  \bibinfo{journal}{Phys. Rev. Lett.} \textbf{\bibinfo{volume}{76}},
  \bibinfo{pages}{730} (\bibinfo{year}{1996}), \eprint{hep-ph/9508373}.

\bibitem[{\citenamefont{Cho and Leibovich}(1996)}]{Cho:1996cg}
\bibinfo{author}{\bibfnamefont{P.~L.} \bibnamefont{Cho}} \bibnamefont{and}
  \bibinfo{author}{\bibfnamefont{A.~K.} \bibnamefont{Leibovich}},
  \bibinfo{journal}{Phys. Rev.} \textbf{\bibinfo{volume}{D54}},
  \bibinfo{pages}{6690} (\bibinfo{year}{1996}), \eprint{hep-ph/9606229}.

\bibitem[{\citenamefont{Bodwin and Petrelli}(2002)}]{Bodwin:2002aa}
\bibinfo{author}{\bibfnamefont{G.~T.} \bibnamefont{Bodwin}} \bibnamefont{and}
  \bibinfo{author}{\bibfnamefont{A.}~\bibnamefont{Petrelli}},
  \bibinfo{journal}{Phys. Rev.} \textbf{\bibinfo{volume}{D66}},
  \bibinfo{pages}{094011} (\bibinfo{year}{2002}), \eprint{hep-ph/0205210}.

\bibitem[{\citenamefont{Braaten and Lee}(2003)}]{Braaten:2002aa}
\bibinfo{author}{\bibfnamefont{E.}~\bibnamefont{Braaten}} \bibnamefont{and}
  \bibinfo{author}{\bibfnamefont{J.}~\bibnamefont{Lee}},
  \bibinfo{journal}{Phys. Rev.} \textbf{\bibinfo{volume}{D67}},
  \bibinfo{pages}{054007} (\bibinfo{year}{2003}), \eprint{hep-ph/0211085}.

\bibitem[{\citenamefont{Bodwin et~al.}(2003{\natexlab{a}})}]{Bodwin:2003aa}
\bibinfo{author}{\bibfnamefont{G.~T.} \bibnamefont{Bodwin}}
  \bibnamefont{et~al.}, \bibinfo{journal}{Phys. Rev.}
  \textbf{\bibinfo{volume}{D67}}, \bibinfo{pages}{054023}
  (\bibinfo{year}{2003}{\natexlab{a}}), \eprint{hep-ph/0205352}.

\bibitem[{\citenamefont{Hagiwara et~al.}(2003)}]{Hagivara:2003aa}
\bibinfo{author}{\bibfnamefont{K.}~\bibnamefont{Hagiwara}}
  \bibnamefont{et~al.}, \bibinfo{journal}{Phys.Lett.}
  \textbf{\bibinfo{volume}{B570}}, \bibinfo{pages}{39} (\bibinfo{year}{2003}),
  \eprint{hep-ph/0305102}.

\bibitem[{\citenamefont{Wang et~al.}(2002)}]{Wang:2002aa}
\bibinfo{author}{\bibfnamefont{P.}~\bibnamefont{Wang}} \bibnamefont{et~al.}
  (\bibinfo{year}{2002}), \eprint{hep-ex/0210062}.

\bibitem[{\citenamefont{Aubert et~al.}(2001)}]{Aubert:2001pd}
\bibinfo{author}{\bibfnamefont{B.}~\bibnamefont{Aubert}} \bibnamefont{et~al.},
  \bibinfo{journal}{Phys.Rev.Lett.} \textbf{\bibinfo{volume}{87}},
  \bibinfo{pages}{162002} (\bibinfo{year}{2001}), \eprint{hep-ex/0106044}.

\bibitem[{\citenamefont{Abe et~al.}(2002{\natexlab{a}})}]{Abe:2001aa}
\bibinfo{author}{\bibfnamefont{K.}~\bibnamefont{Abe}} \bibnamefont{et~al.},
  \bibinfo{journal}{Phys.Rev.Lett.} \textbf{\bibinfo{volume}{88}},
  \bibinfo{pages}{052001} (\bibinfo{year}{2002}{\natexlab{a}}),
  \eprint{hep-ex/0110012}.

\bibitem[{\citenamefont{Kiselev et~al.}(1994)\citenamefont{Kiselev, Likhoded,
  and Shevlyagin}}]{Kiselev:1994pu}
\bibinfo{author}{\bibfnamefont{V.~V.} \bibnamefont{Kiselev}},
  \bibinfo{author}{\bibfnamefont{A.~K.} \bibnamefont{Likhoded}},
  \bibnamefont{and} \bibinfo{author}{\bibfnamefont{M.~V.}
  \bibnamefont{Shevlyagin}}, \bibinfo{journal}{Phys. Lett.}
  \textbf{\bibinfo{volume}{B332}}, \bibinfo{pages}{411} (\bibinfo{year}{1994}),
  \eprint{hep-ph/9408407}.

\bibitem[{\citenamefont{Berezhnoy and Likhoded}(2003)}]{Berezhnoy:2003hz}
\bibinfo{author}{\bibfnamefont{A.~V.} \bibnamefont{Berezhnoy}}
  \bibnamefont{and} \bibinfo{author}{\bibfnamefont{A.~K.}
  \bibnamefont{Likhoded}} (\bibinfo{year}{2003}), \eprint{hep-ph/0303145}.

\bibitem[{\citenamefont{Leibovich et~al.}(2003)}]{Leibovich:2003aa}
\bibinfo{author}{\bibfnamefont{A.}~\bibnamefont{Leibovich}}
  \bibnamefont{et~al.}, \bibinfo{journal}{Phys.Rev.}
  \textbf{\bibinfo{volume}{D68}}, \bibinfo{pages}{094011}
  (\bibinfo{year}{2003}), \eprint{hep-ph/0306139}.

\bibitem[{\citenamefont{Luchinsky}(2003{\natexlab{a}})}]{Luchinsky:2003ej}
\bibinfo{author}{\bibfnamefont{A.~V.} \bibnamefont{Luchinsky}}
  (\bibinfo{year}{2003}{\natexlab{a}}), \eprint{hep-ph/0305253}.

\bibitem[{\citenamefont{Abe et~al.}(2002{\natexlab{b}})}]{Abe:2002aa}
\bibinfo{author}{\bibfnamefont{K.}~\bibnamefont{Abe}} \bibnamefont{et~al.},
  \bibinfo{journal}{Phys.Rev.Lett.} \textbf{\bibinfo{volume}{89}},
  \bibinfo{pages}{142001} (\bibinfo{year}{2002}{\natexlab{b}}),
  \eprint{hep-ex/0205104}.

\bibitem[{\citenamefont{Bodwin et~al.}(2003{\natexlab{b}})\citenamefont{Bodwin,
  Lee, and Braaten}}]{Bodwin:2002fk}
\bibinfo{author}{\bibfnamefont{G.~T.} \bibnamefont{Bodwin}},
  \bibinfo{author}{\bibfnamefont{J.}~\bibnamefont{Lee}}, \bibnamefont{and}
  \bibinfo{author}{\bibfnamefont{E.}~\bibnamefont{Braaten}},
  \bibinfo{journal}{Phys. Rev. Lett.} \textbf{\bibinfo{volume}{90}},
  \bibinfo{pages}{162001} (\bibinfo{year}{2003}{\natexlab{b}}),
  \eprint{hep-ph/0212181}.

\bibitem[{\citenamefont{Luchinsky}(2003{\natexlab{b}})}]{Luchinsky:2003yh}
\bibinfo{author}{\bibfnamefont{A.~V.} \bibnamefont{Luchinsky}}
  (\bibinfo{year}{2003}{\natexlab{b}}), \eprint{hep-ph/0301190}.

\bibitem[{\citenamefont{Groom et~al.}(2000)}]{Groom:2000in}
\bibinfo{author}{\bibfnamefont{D.~E.} \bibnamefont{Groom}} \bibnamefont{et~al.}
  (\bibinfo{collaboration}{Particle Data Group}), \bibinfo{journal}{Eur. Phys.
  J.} \textbf{\bibinfo{volume}{C15}}, \bibinfo{pages}{1}
  (\bibinfo{year}{2000}).

\end{thebibliography}
\end{document}